\documentclass[aps,pre,twocolumn]{revtex4-1}
\usepackage{epsfig,graphicx}
\usepackage{natbib}
\usepackage{ulem}
\usepackage{xspace,xcolor}
 
\usepackage{bbold}
\usepackage{appendix}
\usepackage{amsmath,amsfonts,amssymb}

\newcommand{\w}{\omega}

\begin{document}

\title{
Heat conduction in chains of non-locally coupled harmonic oscillators: mean-field limit}

\author{Lucianno  Defaveri$^1$}
\author{Carlos Olivares$^2$}
\author{Celia Anteneodo$^{1,3}$}
 \affiliation{$^1$Department of Physics, PUC-Rio, Rio de Janeiro, 22451-900 RJ, Brazil}
\affiliation{$^2$Laboratoire de Physique Th\'eorique Ecole Normale Sup\'erieure, France}
\affiliation{$^3$Institute of Science and Technology for Complex Systems,  Brazil}

\begin{abstract} 
We consider  one-dimensional systems of all-to-all harmonically coupled particles with arbitrary masses, subject to two Langevin thermal baths. 
The couplings correspond to the mean-field limit of  long-range interactions. 
Additionally,  the particles can be subject to a harmonic on-site potential to break momentum conservation. 
Using the non-equilibrium Green operator formalism, we  calculate the transmittance, the heat flow and local temperatures, for arbitrary configurations of masses.
For identical masses,  we show analytically that, the heat flux decays  with the system size $N$, as $1/N$, regardless of the conservation or not  of the momentum, and of the introduction or not of a Kac factor. These results describe in good agreement the thermal behavior of systems with small heterogeneity in the masses.
\end{abstract}

\maketitle

\section{Introduction}

Simplified microscopic models, such as classical particle chains in contact with heat baths, have proven useful to grasp  the physics of thermal transport~\cite{BeniniLepriLivi2020review,ReviewLepriLiviPoliti2003,ReviewDhar2008,LepriBook2016}. Specially, the role of conserved quantities in the violation of Fourier's law has been extensively studied so far~\cite{fourier, NarayanRamaswamy2002,Landi2014,LiLiuLiHanggiLi2015,Liu2014PRL,DasDhar2015,Giardina2000}.  
Actually, the interest in one-dimensional models goes beyond  the  higher accessibility from a theoretical approach, as they can also be useful for  understanding the heat conduction anomalies  observed in real systems, such as carbon nanotubes~\cite{nano2008}, silicon  nanowires~\cite{YangZhangLi2010},  molecular chains~\cite{chain1,chain2}, and others~\cite{expChang2016}.
In particular, these  experiments and theories can
lead to new developments based on phonon transport, as thermal diodes~\cite{e1,exp-diode,casati-diode,Tdependence}. 
In this latter context, the range of the interactions 
may be relevant  to increase  rectification~\cite{efficiency1,efficiency2}. 
More generally, sufficiently long-range interactions are worth of investigation as they can bring new physical 
features to a system~\cite{CampaDauxoisRuffoReview2009,LevinPakterRizzatoTelesBenetti2014,Gupta2017Review,RuffoBook}. Among them, let us cite  negative specific heat~\cite{RuffoBook}, ensemble inequivalence~\cite{BarreMukamelRuffo2001}, 
phase transitions even in one-dimensional systems~\cite{Anteneodo2000,CampaGiansantiMoroni2000,anteneodo2004,RochaFilho2011}, slow relaxation and long-lived quasi-stationary states~\cite{Antoni1995,LatoraRapisardaTsallis2001,MukamelRuffoSchreiber2005PRL,MoyanoAnteneodo2006PRE,RochaFilho2014}. 

In the context of heat conduction, 
the range of the interactions has been investigated more recently, mainly through molecular dynamics simulations~\cite{Olivares2016,BagchiTsallis2017,Bagchi2017,Iubini2018,dicintio2019,livi2020,Bagchi2021,Bagchi2017hmf,xiong2020}.
Variants of  Fermi-Pasta-Ulam-Tsingou~\cite{Bagchi2017,Iubini2018,dicintio2019,livi2020,Bagchi2021}, and XY ~\cite{Olivares2016,Bagchi2017hmf,Iubini2018} 
chains, with interactions that decay algebraically with the interparticle distance,  have  been studied. 
But few analytical results exist for long-range systems in this context. 
Among them, let us remark the contribution by Tamaki and Saito~\cite{saito2020}, who considered chains of long-range coupled harmonic oscillators and studied thermal properties 
through the Green-Kubo formula that relates the equilibrium energy current correlation function to the thermal conductivity.  
However, for sufficiently long-range systems, the divergence of the current correlation hampers that calculation. 
The infinite-range limit of a network of harmonic oscillators,  when springs are random,  has been previously tackled  through a random matrix approach~\cite{shapiro2013}. 
In the present work, we consider another variant of the globally coupled harmonic system, with identical couplings, 
and use the non-equilibrium Green function formalism to calculate analytically the heat current $J$, via the transmittance, as a function of the system size $N$. 
In contrast to the well-known case of harmonic   first-neighbor interactions, for which 
the heat current becomes constant for large $N$,  we find that 
for the opposite extreme of infinite-range interactions, the current decays as $1/N$. This result also contrasts with that found when spring disorder is introduced~\cite{shapiro2013}.

 In Sec.~\ref{sec:model}, we describe the model system. Following the non-equilibrium Green function approach, in Sec.~\ref{sec:results}, 
 we calculate
the transmittance, heat flow and local temperature, showing the behavior with system size, for different mass distributions and analytical results are shown for identical masses. 
  Sec.~\ref{sec:final} contains final remarks.

\section{Model}
\label{sec:model}

\begin{figure}[b]
    \centering
    \includegraphics[width=0.5\textwidth]{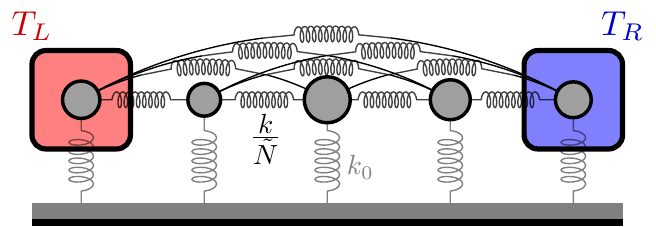}
    \caption{Pictorial representation of the system.}
    \label{fig:picture}
\end{figure}

We consider a system of $N$ globally coupled  harmonic oscillators described by the Hamiltonian  
%
\begin{eqnarray} \nonumber
  \mathcal{H} &=& \sum_{n=1}^N  \frac{p_n^2}{2 m_n  } +\frac{k_0}{2} \sum_{n=1}^N   q_n^2 
+ 
    \frac{k}{ 2\tilde{N} }  
    \sum_{n=1}^N  \sum_{j=1 \atop{j \neq n} }^N   \frac{1}{2}(q_n - q_j)^2, \\
    &&
\label{eq:H}
\end{eqnarray}
where $p_n$ and $q_n$ are respectively the 1D momentum and displacement~\cite{saito2020} 
of the $n$th oscillator with mass $m_n$, while
$k_0$ and $k$ are the stiffness constants of the pinning and the internal interactions respectively and 
 $\tilde{N}$ is a factor that, when setting 
$\tilde{N}=N-1$, it represents
the Kac factor~\cite{Kac1963} warranting  extensivity  in the thermodynamic limit (TL), but  $\tilde{N}=1$ will also be considered,  for comparison with previous literature. 
This system is very similar to that studied by Schmidt et al.~\cite{shapiro2013}, but in that case spring constants are random  and a Kac factor is not used.  The impact of these differences will be commented throughout this paper.  

Notice that the last term of the Hamiltonian can be seen as the infinite-range limit of a chain 
of harmonic oscillators that 
interact with a strength that 
decays with distance between particles. 
In this limit, however,  the spatial order of the chain is lost.  
A schematic representation of the system is given in Fig.\ref{fig:picture}. 

Langevin thermostats are put in contact 
with two of the oscillators. Let's choose the 1st and $N$th ones. The resulting  equations of motion are
 \begin{eqnarray} \label{eq:1}
   m_1\,  \ddot{q}_1 &=&  F_1  
   -\gamma \, \dot{q_1} + \eta_L, \\  \label{eq:n}
   m_n\,  \ddot{q}_n &=& F_n, \hspace{1cm} \mbox{  $n\neq 1,N$}, \\ \label{eq:mu}
   m_N\,  \ddot{q}_{N} &=& F_N  
   -\gamma \,\dot{q}_{N} + \eta_R, 
\label{eq:alphaXYEqMotionGenLang}
\end{eqnarray}
where $\gamma$ is the friction coefficient, $\eta_{L/R}$  independent  fluctuating zero-mean Gaussian  forces, such that
$\langle \eta_{L/R}(t)\,\eta_{L/R}(t') \rangle =2\gamma  T_{L/R}\,\delta(t-t')$, 
$\langle \eta_L(t) \eta_R(t') \rangle = 0$ 
and  the force  over particle $n$ is
\begin{eqnarray} \label{eq:F}
F_n &=& -k_0\, q_n +
 \frac{k}{ \tilde{N} }
\sum_{j=1 \atop{j \neq n}} (q_j - q_n).
\end{eqnarray}

Fourier transforming the 
equations of motion (\ref{eq:1})-(\ref{eq:mu}), 
through the definition
$\hat{x}(\w)=\int_{-\infty}^{\infty} x(t) e^{-i \w t} dt$, 
in matrix form, they become 
\begin{equation} \label{eq:motion-Fourier}
 \hat{Z}(\w) \hat{q}(\w) = \hat{\eta} (\w)  \,,
\end{equation}
where 
$\hat{q}(\w)=(\hat{q}_1(\omega),\ldots,\hat{q}_N(\omega))^T$ is the Fourier-transformed column vector of displacements,  
 $\hat{\eta}(\omega) = ( \eta_L(\w),0,\ldots,0, \eta_R(\w))^T$ 
 the noise column vector, 
 and    $N\times N$ matrix $\hat{Z}(\w) $ has the symmetric form
\begin{equation}
\hat{Z}(\omega) = 
\left(
\begin{array}{ccccc}
a_1+c   &b   &\cdots   &b   &b\\
b       &a_2 &b   &\cdots   &b\\
\vdots &    &\ddots      &        &\vdots \\
b      &\cdots    &b &a_{N-1}  &b\\
b   & b    &\cdots    &b &a_N+c\\
\end{array}
\right)   \,,
\label{eq:Z}
\end{equation}
where
\begin{eqnarray}   \label{eq:a}
 a_n  &=&  \frac{N-1}{\tilde{N}} \, k +k_0-m_n\,\omega^2, \\ \label{eq:bn}
 b  &=& -\frac{k}{\tilde{N}} \,, \\  \label{eq:c}
 c &=&  i\w \gamma\,.
 \end{eqnarray}
 The inverse matrix $\hat{G}=  \hat{Z}^{-1}$
is the Green operator that provides the solution of the system of equations  (\ref{eq:1})-(\ref{eq:mu}). 

\section{Results}
 \label{sec:results}
 
  The elements of the matrix  $\hat{G}=\hat{Z}^{-1}$ can be obtained as
 \begin{equation}
 \hat{G}_{ij}  = \hat{Z}^{-1}_{ij} 
 =\frac{ (-1)^{i+j}\hat{M}_{ij} }{{\rm det}(\hat{Z})},
\end{equation}
where ${\rm det}(\hat{Z})$ is the determinant of the matrix $\hat{Z}$, and $\hat{M}_{ij}$ is  the ($i,j$) minor (i.e.,  the determinant of the sub-matrix that results from the elimination of the $i$th row and $j$th column of $\hat{Z}$). 
 Derivations are essentially done through 	Laplace expansion of a determinant by  minors. 
 
For the modulus of the ($i,j$) minor, we straightforwardly obtain 
 \begin{eqnarray} \label{eq:Mij}
    |\hat{M}_{ij}| = \left|  \frac{b}{A_i A_j}\prod_{n=1}^N A_n \right| \,,
\end{eqnarray}
for $i\neq j$, where we have defined
\begin{eqnarray} \nonumber
    A_i = a_i-b+c(\delta_{i1}+\delta_{iN}).
\end{eqnarray}
For $i=j$, the minor corresponds to the determinant of the matrix $\hat{Z}$ of reduced order.

The modulus of the determinant of the 
$N\times N$ matrix $\hat{Z}$  is
\begin{eqnarray} \label{eq:detZ}
    |{\rm det}(\hat{Z})| = 
    \left| \left(  1 + \sum_{j=1}^N \frac{b}{A_{j}}  \right)   \prod_{n=1}^N A_n  
    \right|\,.
\end{eqnarray}
Then, for $i\neq j$, 
\begin{eqnarray} \label{eq:Gij}
    |\hat{G}_{ij}| =   \left|\frac{A_i A_j}{b} \left(\displaystyle 1 + \sum_{n=1}^N \frac{b}{A_n} \right)\right|^{-1} \,.
\end{eqnarray}

 In the next subsections, we will use the Green operator $\hat{G}$ to find the  heat flux and local temperature. Their mathematical expressions in terms of the elements of $\hat{G}$ are formally  the same previously derived in the literature for first-neighbor interactions (see for instance~\cite{DharPRL2001,Dhar2015Review}), which are actually valid for any interaction network. In our case, it is given by Eq.~(\ref{eq:Z}), where all  off-diagonal elements are non-null, due to the all-to-all interactions, in contrast to the tri-diagonal first-neighbor case.

\subsection{Transmittance and heat flux}
\label{sec:flux}

In a long-range system, with all-to-all interactions, the bulk particle can receive heat through many channels, but we can calculate, without ambiguity, the fluxes that enter and leave the system~\cite{Iubini2018}, respectively from the left bath to the first particle or from the rightmost particle to the right bath, that must coincide under stationary conditions, i.e., 
\begin{equation} \label{eq:JJ}
J  = \langle (\eta_L -\gamma\dot{q}_1)
 \dot{q}_1 \rangle  = -\langle (\eta_R -\gamma\dot{q}_N)
 \dot{q}_N \rangle\,, 
 \end{equation}
which has the form
\begin{eqnarray} \label{eq:JT}
 J    &=&\frac{T_L - T_R}{4 \pi} \int_{-\infty}^{\infty} \mathcal{T}(\omega) d\omega \,,
\end{eqnarray}
where $\mathcal{T}(\omega)$ 
is the transmission coefficient
\begin{equation}
  \mathcal{T}(\omega)  =  4 \gamma^2 \w^2 |\hat{G}_{1N} (\omega)|^2\,,   
\label{eq:Tdef}
\end{equation}
which  depends on the bath properties (given only by $\gamma$ in the case of our choice of baths) and on the system, via the element $\hat{G}_{1N}$, which can be obtained from Eq.~(\ref{eq:Gij}).
%

Let us consider  the particular 
case of identical masses, $m_n=m$, for all $n$. 
As we will see,  this case yields normal modes uncoupled to the baths, however, the analytical results still apply in the limit of small heterogeneity in the masses,  enough to recover the coupling with the baths.
From Eq.~(\ref{eq:Gij}),  we obtain 
\begin{equation}   
| \hat{G}_{1N}| = \frac{| b(a-b)|}{|a-b+c|\, |\left(a-b+Nb\right)\left(a-b+c\right)-2c b|}.
 \\ \label{eq:G1n}
\end{equation}

Using the definitions of  $a$, $b$, $c$ given by Eqs.~(\ref{eq:a}), (\ref{eq:bn}) and (\ref{eq:c}), we have
\begin{eqnarray} \nonumber
 \mathcal{T}(\w) &=&   4 \gamma^2 w^2  |\hat{G}_{1N}|^2 = 4 \gamma^2 w^2 \hat{G}_{1N}(\w)\hat{G}_{1N}(-\w) \\[3mm] 
&=& \frac{   4 \gamma^2 \w^2 k^2( \frac{Nk}{\tilde{N}}+k_0-m\w^2)^2}{ f(\w) \Delta(\w)\tilde{N}^2 } \,, 
\label{eq:Tgen}
\end{eqnarray}
where 
\begin{eqnarray} \label{eq:f}
   f(\w) &=& \left(\frac{Nk}{\tilde{N}}+k_0-m\w^2 \right)^2+\gamma^2\w^2\,, \\ \nonumber
   \Delta(\w) &=&  \gamma^2\w^2\left( \frac{2k}{\tilde{N}}+k_0-m\w^2  \right)^2  \\ \label{eq:delta}
&+& \Bigl(
 k_0-m\w^2 \Bigr)^2   \Bigl(\frac{Nk}{\tilde{N}}   + k_0-m\w^2 
\Bigr)^2.
\end{eqnarray}

Let us remark  that Eqs.~(\ref{eq:Tgen})-(\ref{eq:delta}) are valid for any $\tilde{N}$. We will analyze separately the momentum conserving ($k_0=0$) and non-conserving ($k_0>0$) cases.   
In the first case, mathematical expressions  are common to both values of $\tilde{N}$, in the second, some expressions will be split for each value of  $\tilde{N}$.

\begin{figure}[b!]
\hspace*{-0.2cm}
\includegraphics[width=0.5\textwidth]{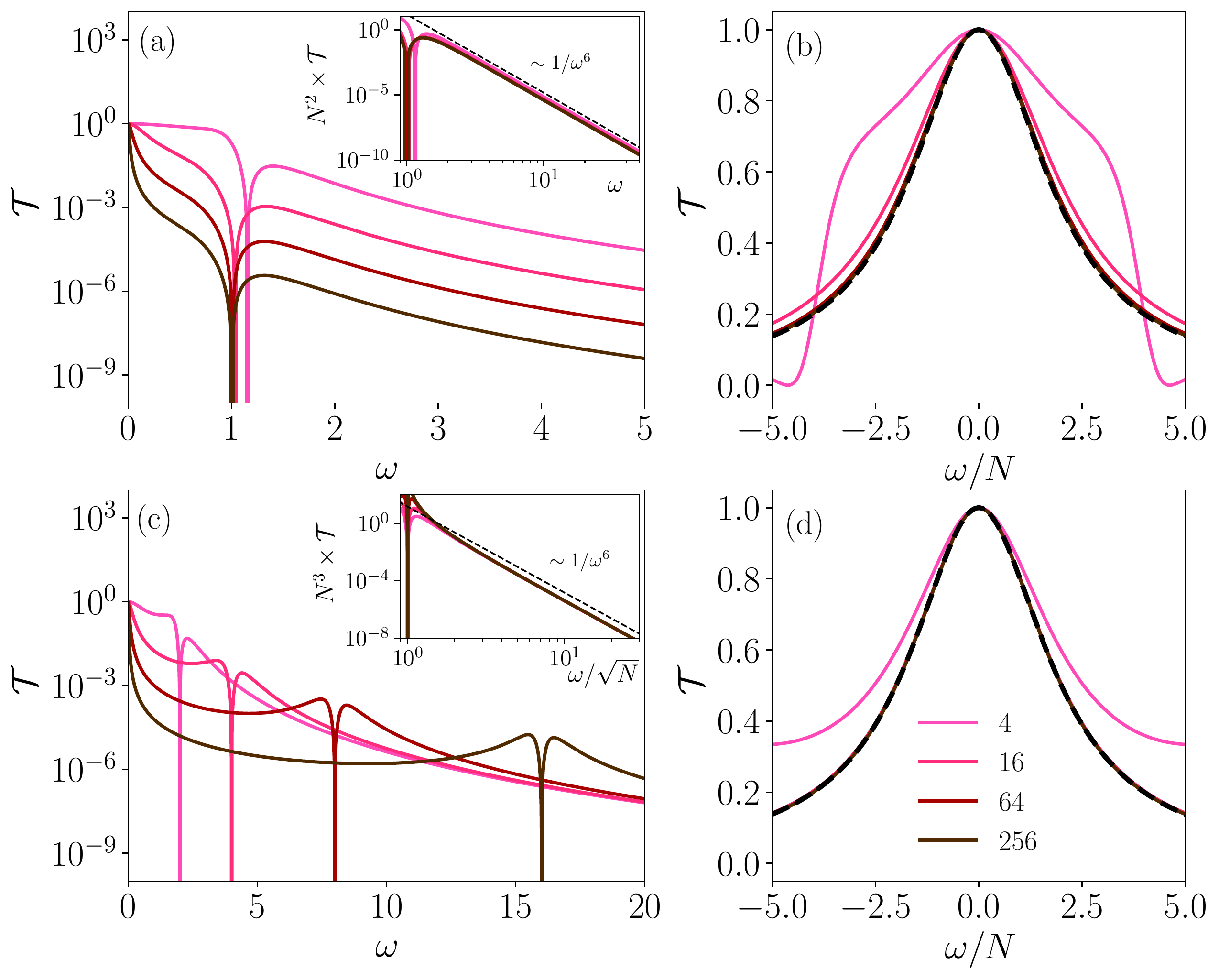}
 \caption{Transmittance $\mathcal{T}(\w)$ in the momentum conserving case  (only the positive abscissa is displayed, having in mind that $\mathcal{T}$ is an even function),   
 for different values of   $N$ indicated in the legend, 
 using Eq.~(\ref{eq:T0}). 
  In all cases we used  
  $m=k=\gamma=1$ and $k_0=0$. 
  In panels (a)-(b), $\tilde{N}=N-1$ and in panels (c)-(d) $\tilde{N}= 1$.  
In (a)-(c), the insets highlight the scaling of the transmittance for large $\w$. 
In (b) and (d), we focus on the main peak, and the black lines correspond to the Lorentzian 
 approximation given by Eq.~(\ref{eq:T0app0}), using the respective values of $\tilde{N}$. 
 }
 \label{fig:trans-conserving}
\end{figure} 

(i) When $k_0=0$, Eq.~(\ref{eq:Tgen}) reduces to
\begin{eqnarray}
\mathcal{T}(\w) &=&    \frac{   4 \gamma^2 k^2( \frac{Nk}{\tilde{N}}-m\w^2)^2\frac{1}{\tilde{N}^2}}{ 
 f(\w) [m^2\w^2 f(\w) 
 - 4 \gamma^2 k(\frac{m \w^2}{\tilde{N}}-\frac{k}{\tilde{N}^2})] }\,, \;\;\;\;\;\;\;\label{eq:T0}
\end{eqnarray}
where now $f(\w)= (\frac{Nk}{\tilde{N}}-m\w^2)^2+\gamma^2\w^2$. 
$\mathcal{T}(\w)$ has an absolute maximum at $\w=0$. 
For   
$N^2/\tilde{N}\gg  4\gamma^2/(mk)$,   only the maximum at $\w=0$ dominates  (additional  maxima with $\mathcal{T}<1$ can emerge for small $N$). 
Therefore, for large enough $N$, and 
$\w^2 \ll N k/(\tilde{N}m)$,
Eq.~(\ref{eq:T0}) approaches  
\begin{eqnarray}
\mathcal{T}(\w)    
&\simeq& \frac{   1}{ 
1+ N^2\,\bigl(\frac{m}{2\gamma}\bigr)^2 \w ^2  }\,, \;\;\;\;\;\;\;\label{eq:T0app0}
\end{eqnarray}
which is a Lorentzian with width that  scales as $1/N$. 
This Lorentzian peak is associated to the  complex conjugate pair of poles that get closer to the real axis when  the dissipation parameter $\gamma/m$ decreases.  
Eq.~(\ref{eq:T0app0}) holds both with or without Kac factor, and it is compared to exact results 
 in   Fig.~\ref{fig:trans-conserving}.

 Moreover,  for large $\w^2\gg Nk/{(\tilde{N} m)}$, the transmittance decays as $\mathcal{T} \sim 1/(w^6 \tilde{N}^2)$. 
 This is  depicted in the insets of Fig.~\ref{fig:trans-conserving} for the 
 respective values of  $\tilde{N}$. 
 Hence, the  integral in Eq.~(\ref{eq:JT}) is dominated by the Lorentzian peak described by Eq.~(\ref{eq:T0app0}),  leading to
\begin{eqnarray} \label{eq:J-large-N}
J/\Delta T 
&\simeq& \frac{ \gamma/(2m)}{N}. 
\end{eqnarray}
  
Furthermore, let us comment that, when the Kac factor is introduced, the behavior 
 $ \mathcal{T}(\w)  \sim 1/N^2$  is evident  
 for any $\w$
except in the global maxima where 
 $\mathcal{T}=1$ and in the zeros where  $\mathcal{T}=0$. 
If the Kac factor is eliminated, the same law does not hold for any frequency but still holds in the dominant region.

\begin{figure}[h!]
\includegraphics[width=0.45\textwidth]{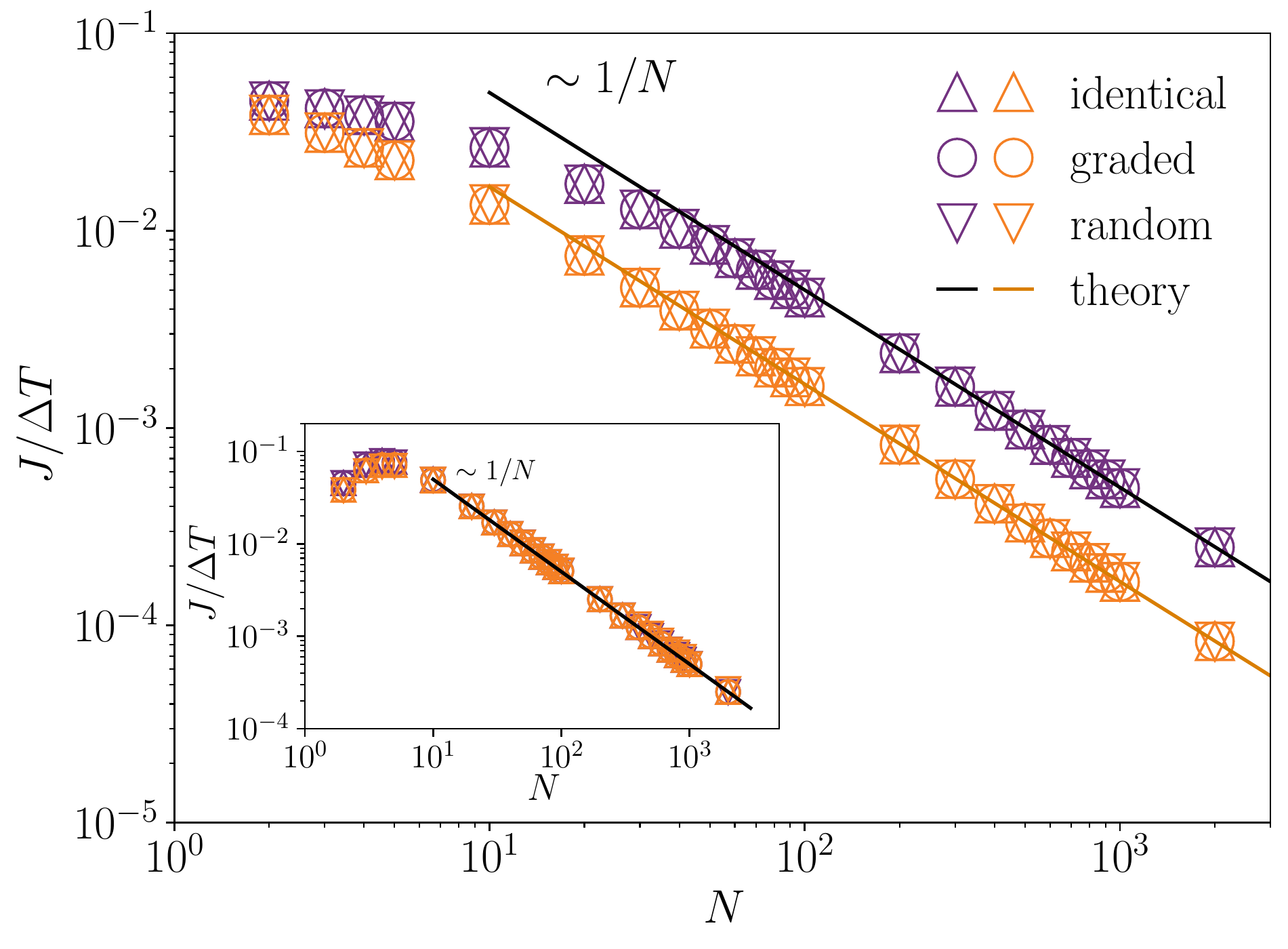}
 \caption{Heat flux vs. system size. 
  The symbols correspond to the numerical integration of Eq.~(\ref{eq:Tdef}), while the solid lines correspond  to the theoretical approximations valid for large $N$:   Eqs.~(\ref{eq:J-large-N}), for $k_0=0$ and$|$or $\tilde{N}=1$, and Eq.~(\ref{eq:J-k0-large-N}), otherwise.
 In all cases
 $\gamma=1$, $k=0.1$, with $k_0 = 0$ (dark lilac) or $k_0 = 0.02$ (light orange). 
 In the main plot, we used $\tilde{N}=N-1$, while in the inset $\tilde{N}=1$, for the same values of the parameters. 
 Besides identical masses $ {m}=1$, we also considered slightly graded and random distributions, with amplitude $\delta = 0.1$. 
}
 \label{fig:flux}
\end{figure}

(ii) In the  case $k_0 > 0$ (non-conserving), 
 Eq.~(\ref{eq:T0}) takes the maximal value  1 at  frequencies $\pm \w_c$, given by
\begin{eqnarray}  \label{eq:wc}
    \w_c = \left\{
    \begin{array}{ll}
         \w_0 +\frac{\w_0}{N}\frac{k\gamma^2/m }{ k^2    +\w_0^2\gamma^2  } + O(  N^{-2} ), &\mbox{if $\tilde{N}=N-1$},\\
         \w_0 + O(  N^{-2} ),&\mbox{if $\tilde{N}=1$},
    \end{array} \right.  \;\;\;\;\;
\end{eqnarray}
with $\w_0 = \sqrt{k_0/m}$.  These resonance frequencies can be obtained by solving ${\cal T}(\omega_c) = 1$, up to the first order of $1/N$. 
Other peaks are avoided for large enough $N$, namely verifying 
\begin{eqnarray}  \label{eq:N0}
    \begin{array}{ll}
         N \gg\frac{4k\gamma^2}{m}\frac{k^2 -\w_0^2\gamma^2}{(k^2  +\w_0^2 \gamma^2)^2}, &\mbox{if $\tilde{N}=N-1$},\\[2mm]
         N^2 \gg \frac{4\gamma^2}{mk}-\frac{2\w_0^2\gamma^2}{k^2},&\mbox{if $\tilde{N}=1$},
    \end{array}.
\end{eqnarray}
In such case, the transmittance tends to the superposition of two Lorentzian peaks that narrow with increasing $N$ (see   Fig.~\ref{fig:trans-conserving}), as
\begin{eqnarray}
\mathcal{T}(\w)    
&\simeq& \sum_{\Omega=\pm \w_c}\frac{   1}{ 
1+ N^2 {\cal A}^2(\w-\Omega)^2 } \,,  \label{eq:T0app}
\end{eqnarray}
where
\begin{eqnarray} \label{eq:A}
   {\cal A} = \left\{ \begin{array}{ll}
   \frac{ m}{ \gamma} \bigl( 1+ 
   \frac{  \w_0^2\gamma^2}{  k^2}\bigr), &\mbox{if $\tilde{N} = N-1$}   \\
    \frac{ m}{ \gamma}, & \mbox{if $\tilde{N} = 1$}.
   \end{array} \right.
\end{eqnarray}
The frequencies that significantly contribute to the transmission are those around $\pm\w_c$, with bandwiths decreasing as $1/N$, which   signals a localization \cite{AshPRB2020, CaneArxiv2021}. 
Hence, also in this case 
 \begin{eqnarray}
J/\Delta T 
&\simeq& \frac{1}{ {2\cal A}N} .
\label{eq:J-k0-large-N}
\end{eqnarray}
This expression recovers Eq.~(\ref{eq:J-large-N}) when $k_0=0$, and shows why when $\tilde{N}=1$ the result does not depend on $k_0$, as observed in Fig.~\ref{fig:flux}.

\begin{figure}[h]
\hspace*{-0.2cm}
\includegraphics[width=0.5\textwidth]{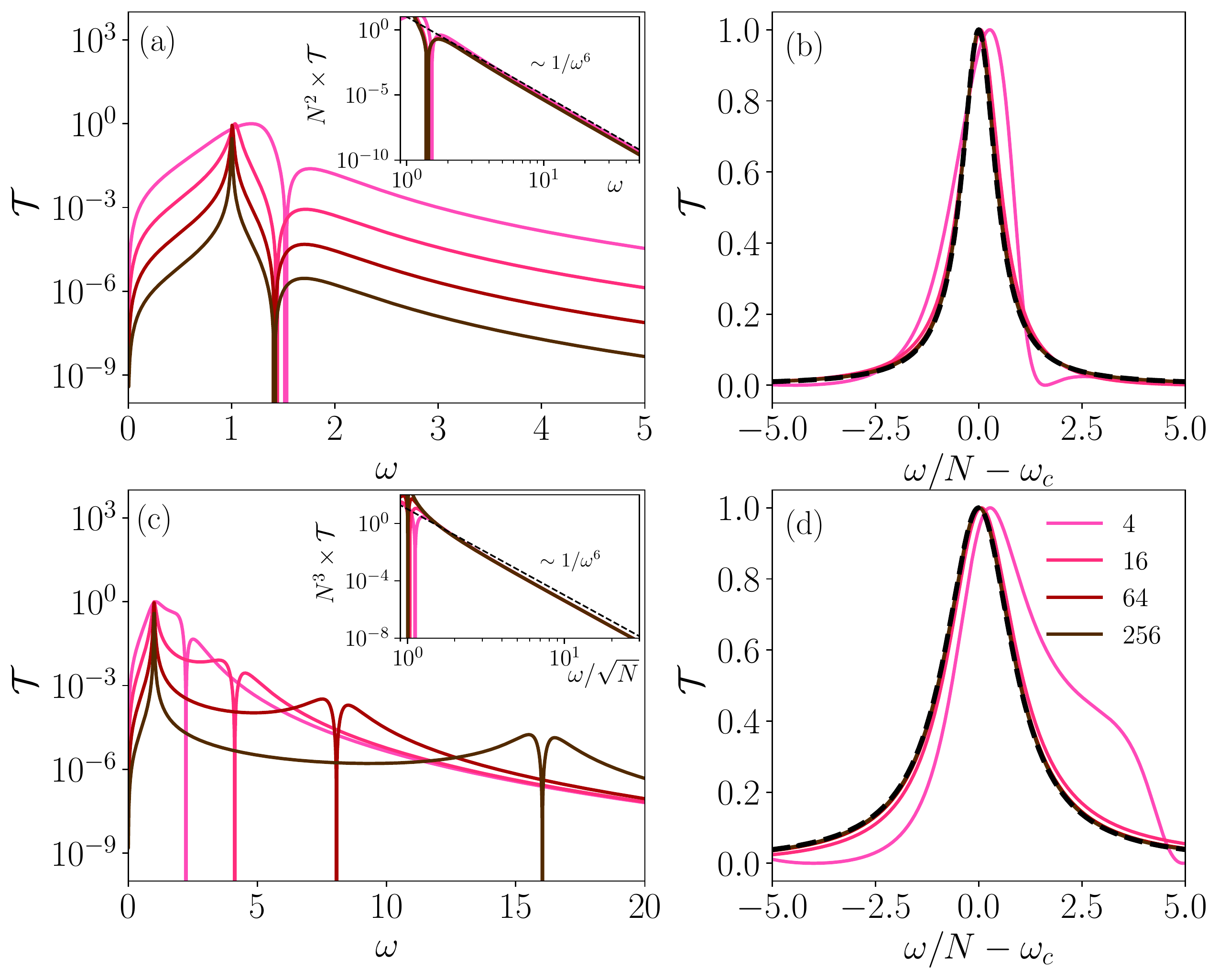}
 \caption{Transmittance $\mathcal{T}(\w)$ in the momentum non-conserving case with $k_0=1$. The remaining parameters are the same used in Fig.~\ref{fig:trans-nonconserving}. 
For the Lorentzian approximation,   Eqs.~(\ref{eq:T0app}) and (\ref{eq:A}) were used and $\omega_c$ is given by Eq.\,(\ref{eq:wc}).  
 }
 \label{fig:trans-nonconserving}
\end{figure} 
In conclusion, the flux decays as $1/N$. 
This result does not depend  on the existence of pinning ($k_0\neq 0$) or  on the introduction of a Kac factor.  
Such picture still holds, when introducing certain degree of heterogeneity. 
These effects are all illustrated  in  Fig.~\ref{fig:flux}, where besides the case of identical masses developed analytically, we included numerical results, by integrating Eq.~(\ref{eq:Tdef}) for other configurations of masses, with variations of small amplitude $\delta  \ll 1$ around the average mass,  namely:  
(i) graded  masses, varying linearly between $ {m}-\delta$ and $ {m}+\delta$, 
that is, following the rule $m_n=m-\delta + 2\delta  (n-1)/(N-1)$, and 
(ii) random masses, uniformly distributed in $[m-\delta,m+\delta]$. 
It is remarkable that, for small deviations from the average mass, the analytical expressions for the heat current, obtained for identical masses, still hold.

\subsection{Local temperatures}
\label{sec:temp}
 
The local temperature $T_n$, associated to the equilibrium position of particle $n$,   is defined as twice its mean kinetic energy. According to the Green function formalism, 
we have
\begin{eqnarray} \nonumber
 && T_n = m_n\langle (\dot{q}_n)^2\rangle = \\ \label{eq:Tl}
 && 2\gamma\,m_n
 \int\limits_{-\infty}^{\infty} \frac{d\w}{2\pi}  \w^2 \biggl[ 
              T_L |\hat{G}_{n1}(\w)|^2   
            + T_R  |\hat{G}_{nN}(\w)|^2  
         \biggr]\,.
\end{eqnarray}

The determinant and  minors required to obtain the elements $\hat{G}_{n1}$ and $\hat{G}_{nN}$ of the Green operator were already defined in  Eqs.~(\ref{eq:detZ}) and ~(\ref{eq:Mij}), respectively. 

\begin{figure}[h!]
\includegraphics[width=0.48\textwidth]{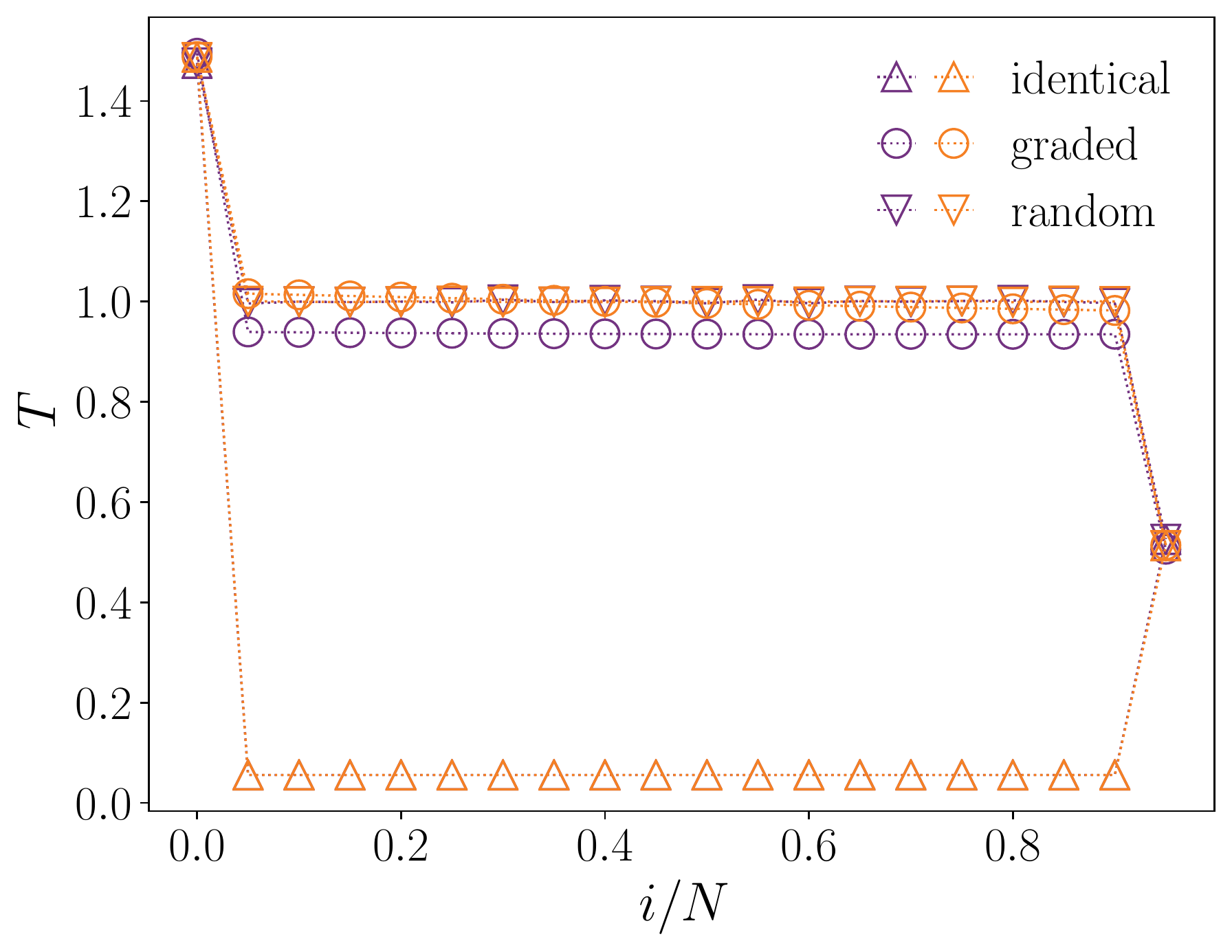}
 \caption{Local temperature,  
for different configurations of masses indicated in the legend. In all  cases, the average mass is $m=1$. 
For random and graded  masses, the amplitude is $\delta=0.1$ . 
 We used 
$k=\gamma=1$, $T_L = 1.5$, $T_R = 0.5$ and $N=20$, 
 with $k_0   = 0$ (light orange) or $k_0 = 1$ (dark lilac). 
 Symbols were obtained by performing the integration in  Eq.~(\ref{eq:Tl})  numerically. 
 Lines are a guide to the eye.
  The results were obtained for $\tilde{N}=N-1$ but are essentially the same for $\tilde{N}=1$. 
 }
 \label{fig:profile}
\end{figure}

In Fig.~\ref{fig:profile}, we show the   local temperatures as a function of the particle index, corresponding to the cases shown in Fig.~\ref{fig:flux}. 
The local temperature of the bulk is always nearly constant but for the case of identical masses, the bulk does not thermalize. 
The bulk will always be comprised of  $N-2$ particles, and the $N-3$ coordinates of relative motion of these bulk particles will be normal modes of the system~\cite{shapiro2013}. 
These modes are not coupled with the heat baths, a consequence of the symmetry of the bulk particles in the system.
If the symmetry is suitably broken, 
even in a minimal way, all modes become coupled.

For instance, for the cases with graded or random masses shown in Fig.~\ref{fig:profile},  the thermal coupling is reestablished. Alternating masses (not shown), however, will not be sufficient to  establish a thermal connection.

  Let us comment that for a graded system with larger amplitude of masses, the temperature ``profile'' will adopt a tilted shape. However, this is artificial in the infinite-range case, where spatial order of the bulk is lost and the transport between baths does not depend on the choice of which are the two particles immersed in the baths.  

\section{Final comments}
\label{sec:final}
 
We obtained exact expressions for the heat current and local temperature for systems of particles with arbitrary masses, coupled through a mean-field network of harmonic interactions. 

We can conclude that in the thermodynamic limit, 
the mean-field flux behaves as  $J\sim 1/N$ and hence  $\kappa=JN/\Delta T$ is asymptotically constant. 
This law is robust against the existence of pining or  the introduction of  a Kac factor. 
We verified that the same scaling also holds if with include a fixed boundary condition (not shown).
 The observed scaling can be associated to the fact that the transmittance is dominated by one (conserving case) or two (non-conserving case) Lorentzian peaks with width that decays as $1/N$. 
Let us note that non-overlapping Lorentzian peaks were also observed in disordered harmonic chains with nearest neighbor interactions, in the weak and strong coupling limits~\cite{AshPRB2020}. 

 For small deviations from the average mass, the analytical expressions for identical masses still hold.
 Differently, the $1/N$ law can be altered by disorder in the couplings, if the Kac factor is not used, in which case the current becomes constant for large  $N$~\cite{shapiro2013}. 
This is observed when the Kac factor is not used, i.e., $\tilde N=1$. 
In fact, in such case, a band appears in the transmittance whose integral does not depend on $N$, and hence dominates over the contribution of the peak that we observe in the absence of stiffness disorder, even for small amplitude of the fluctuations. We also noticed that when the Kac factor is introduced the contribution of this band decreases as $1/N$, hence the heat current too.

Regarding the local temperatures,   similarly to the nearest-neighbors case,  the bulk temperature is  nearly uniform, but the average of the baths is attained only for some  scattering component even if perturbative.

For comparison, let us recall that, 
for nearest neighbors, mass-gradient harmonic chains~\cite{reich2013,massgradient2007}, as well as identical masses,  produce  a constant current. 
However, in (stiffness) disordered chains, with nearest-neighbor interactions, it has been reported that, when disorder has a 
heavy-tailed  distribution, the conductance 
scales as $1/N$~\cite{AshPRB2020}, consistent with Fourier's law.  

Let us also remark that 
since the system is harmonic,   it can be decomposed in $N$ normal modes, then the energy is always extensive, but, for all-to-all interactions without Kac factor, the entropy  is not extensive, and  frequency grows with $\sqrt{N}$. 
If we eliminate the Kac factor by setting $\tilde{N}=1$ in Eq.~(\ref{eq:H}), hence in Eq.~(\ref{eq:bn}),  
we remark that Eq.~(\ref{eq:T0app0}) remains valid, as well as Eq.~(\ref{eq:T0app}), implying that the results for the scaling of the heat current are not altered by using or not the Kac factor in the mean-field model.

It is interesting, to note that for long-range interacting chains, 
with non-harmonic potentials (FPUT, XY), 
the same scaling of the flux between thermal sources $J\sim 1/N$ was observed
~\cite{Iubini2018}, which may be associated to 
harmonic behavior in the limit of low enough temperature.

 As a natural extension of this work, it would be interesting to obtain analytical results for finite range of the interactions, decaying algebraically with the interparticle distance. 
 In this case the chain order would be recovered and the transition between the first-neighbor and mean-field limits accessed.

{\bf Acknowledgments:}  
C. O. gratefully acknowledges enlightening discussions with A. Dhar and S. Lepri. 
C.A.   acknowledges partial financial support from Brazilian agencies CNPq and FAPERJ.  CAPES (code 001) is also acknowledged.


\begin{thebibliography}{99}

  

\bibitem{ReviewLepriLiviPoliti2003}
S. Lepri, R. Livi,  A. Politi, 
{Phys. Rep.}  377, 1  (2003).

\bibitem{ReviewDhar2008}
A. Dhar, 
Adv. in Physics   57, 457  (2008).

 

\bibitem{LepriBook2016}
S. Lepri, R. Livi, A. Politi in  
{\it Thermal Transport in Low Dimensions: From Statistical Physics to Nanoscale Heat Transfer} Chap. 1, 
S. Lepri, editor (Springer, 2016).
%
 
 
 \bibitem{BeniniLepriLivi2020review}
G. Benenti, S. Lepri, R. Livi,
Front. Phys., vol 8, 292 (2020). 


\bibitem{fourier}
F. Bonetto, J.L. Lebowitz, L. Rey-Bellet,
Math. Phys. 2000, 128–150, (2000).
 
 \bibitem{NarayanRamaswamy2002}
O. Narayan, S. Ramaswamy, 
Phys.  Rev.  Lett.  89, 200601 (2002).

  
\bibitem{Landi2014}
G.~T. Landi, M.~J. de~Oliveira, 
Phys.  Rev.  E  89, 022105 (2014).

\bibitem{LiLiuLiHanggiLi2015}
Y. Li, S. Liu, N. Li, P. H\"anggi, B. Li, 
NJP 17, 043064 (2015).
 
  
\bibitem{Liu2014PRL}
S.~Liu, P.~H\"anggi, N.~Li, J.~Ren, B.~Li, 
Phys.  Rev.  L  112,  040601 (2014).
 
  
 
\bibitem{DasDhar2015}
S.~G. Das, A. Dhar, 
arxiv.org/pdf/1411.5247



\bibitem{Giardina2000}
C.~Giardin\'a, R.~Livi, A.~Politi,  M.~Vassalli, 
Phys.  Rev.  L  84, 2144--2147 (2000).


 
\bibitem{nano2008} 
C.W. Chang, D. Okawa, H. Garcia, A. Majumdar, A. Zettl,
Phys.  Rev.  L  101, 075903 (2008).

\bibitem{YangZhangLi2010}
N. Yang, G. Zhang, B. Li, 
 Nano Today  5, 85 (2010).

 
\bibitem{chain1}
Z. Wang et al., 
Science  317 (5839), 787–790 (2007).
 

\bibitem{chain2} 
T.~Meier, F.~Menges, P.~Nirmalraj, H.~H\"olscher, H.~Riel,  B.~Gotsmann,
Phys.  Rev.  L  113, 060801 (2014).


 
\bibitem{expChang2016}
C.-W. Chang, 
{\it Experimental Probing of Non-Fourier Thermal Conductors},
  pp.~305--338.
(Springer, 2016).


\bibitem{e1}
N. Li, J. Ren, L. Wang, G. Zhang, P. H\"anggi, B. Li, 
 Rev. Mod. Phys. 84, 1045  (2012).
 
 \bibitem{exp-diode} 
C.W. Chang, D. Okawa,  A. Majumdar, A. Zettl,
Science  314(5802), 1121–1124 (2006).


\bibitem{casati-diode}
B. Li, L. Wang, G. Casati, 
Phys.  Rev.  L  93, 184301 (2004).



\bibitem{Tdependence}
E. Pereira, 
Phys.  Rev.  E  96, 012114 (2017). 

\bibitem{efficiency1}
E. Pereira, R.R. \'Avila, 
Phys.  Rev.  E   88, 032139 (2013).

\bibitem{efficiency2}
S. Chen, E. Pereira, G. Casati, 
EPL  111, 30004 (2015).


 

 

\bibitem{CampaDauxoisRuffoReview2009}
A.~Campa, T.~Dauxois,  S.~Ruffo, 
Phys.Rep.  480,  57 (2009).
 
 \bibitem{LevinPakterRizzatoTelesBenetti2014}
Y.~Levin, R.~Pakter, F.~B. Rizzato, T.~N. Teles,   F.~P. Benetti,
Phys.Rep.   535,  1 (2014).

\bibitem{Gupta2017Review}
S.~Gupta, S.~Ruffo, 
 IJMP A 32, 1741018 (2017).  
  
\bibitem{RuffoBook}
T. Dauxois, S., E. Arimondo, M. Wilkens, Eds., {\it  Dynamics and Thermodynamics of Systems with Long Range Interactions}
 (Springer, 2002).


  
\bibitem{BarreMukamelRuffo2001}
J.~Barr\'e, D.~Mukamel,  S.~Ruffo, 
Phys.  Rev.  L  87,  030601 (2001).

 
   
\bibitem{Anteneodo2000}
F. A. Tamarit, C. Anteneodo,
Phys.  Rev.  L   84,  208 (2000).
  
  
  
\bibitem{CampaGiansantiMoroni2000}
A.~Campa, A.~Giansanti,  D.~Moroni, 
Phys.  Rev.  E   62,   303 (2000).
 
  
 \bibitem{anteneodo2004} 
 C. Anteneodo, 
 Physica A 342, 112 (2004).
 
  
 \bibitem{RochaFilho2011}
T.~M. Rocha~Filho, M.~A. Amato, B.~A. Mello,   A.~Figueiredo, 
Phys.  Rev.  E  84,  041121 (2011).


  
\bibitem{Antoni1995}
M.~Antoni, S.~Ruffo, 
 Phys.  Rev.  E   52,  2361 (1995).
 


 
\bibitem{LatoraRapisardaTsallis2001}
V.~Latora, A.~Rapisarda,  C.~Tsallis, 
 Phys.  Rev.  E  64,  056134 (2001).
  
  
 
\bibitem{MukamelRuffoSchreiber2005PRL}
D.~Mukamel, S.~Ruffo,  N.~Schreiber, 
Phys.  Rev.  L  95,  240604 (2005). 

\bibitem{MoyanoAnteneodo2006PRE}
L.~G. Moyano, C.~Anteneodo, 
 Phys.  Rev.  E   74,  021118 (2006).
 
\bibitem{RochaFilho2014}
T.~M. Rocha~Filho, M.~A. Amato, A.~E. Santana, A.~Figueiredo,  J.~R. Steiner, 
Phys.  Rev.  E  89, 032116 (2014). 

 

\bibitem{Olivares2016} 
C. Olivares, C. Anteneodo, 
Phys. Rev. E 94, 042117 (2016).
 

\bibitem{BagchiTsallis2017}
D.~Bagchi, C.~Tsallis, 
  Phys. Lett. A 381, 1123 (2017).
  
  
\bibitem{Bagchi2017}
D.~Bagchi, 
Phys. Rev. E 95,  032102 (2017).


  
  \bibitem{Iubini2018}
S. Iubini, P. Di Cintio, S.Lepri, R. Livi,  L. Casetti,
Phys.  Rev.  E  97, 032102 (2018)

\bibitem{dicintio2019}
P. Di Cintio, S. Iubini, S. Lepri and R Livi,
J. Phys. A: Math. Theor. 52, 274001 (2019).



\bibitem{livi2020}
R. Livi, 
{\it  Heat transport in one dimension}, 
J. Stat. Mech. 034001 (2020).


\bibitem{Bagchi2017hmf}
D.~Bagchi, 
Phys. Rev. E 96, 042121 (2017).



\bibitem{xiong2020}
J. Wang, S. V. Dmitriev, and D. Xiong,  
Phys. Rev. Res. 2, 013179 (2020). 

\bibitem{Bagchi2021}
D.~Bagchi,
arXiv arXiv:2108.03424v1.

\bibitem{saito2020}
S. Tamaki, K. Saito,
Phys. Rev. E 101, 042118 (2020).

\bibitem{shapiro2013}
M. Schmidt, T. Kottos, B. Shapiro, 
Phys. Rev. E 88, 022126 (2013).

\bibitem{Kac1963}
M.~Kac, G.~E. Uhlenbeck,   P.~C. Hemmer,  
J.  Math. Phys. 4, 216 (1963).

 
\bibitem{DharPRL2001}
A. Dhar, 
Phys.  Rev.  L  86, 5882  (2001).

\bibitem{Dhar2015Review}
A.~Dhar, R.~Dandekar, 
Physica A  418,  49 (2015).
 


 \bibitem{AshPRB2020}
B. Ash, A. Amir, Y. Bar-Sinai, Y. Oreg, 
and Y. Imry, 
Phys. Rev. B 101, 121403(R) (2020).


\bibitem{CaneArxiv2021}
G. Cane, J. Majeed Bhat, A. Dhar, C. Bernardin, arXiv:2107.06827 


 \bibitem{reich2013}
 K. V. Reich,
Phys. Rev. E 87, 052109 (2013).

\bibitem{massgradient2007}
N. Yang, N. Li, L. Wang, B. Li,
Phys. Rev. B 76, 020301 (R) (2007).





\end{thebibliography}
\end{document}